\documentclass[aps,prl,preprint,onecolumn,superscriptaddress,longbibliography]{revtex4-1}
\usepackage{graphicx}
\usepackage{rotating}
\usepackage{color}
\usepackage{soul}

\begin{document}

\title{Experimental evidence of the failure of Jarzynski equality in active baths}

\author{Aykut Argun}
\affiliation{Soft Matter Lab, Department of Physics, Bilkent University, Cankaya, 06800 Ankara, Turkey}

\author{Ali-Reza Moradi}
\affiliation{Soft Matter Lab, Department of Physics, Bilkent University, Cankaya, 06800 Ankara, Turkey}
\affiliation{Department of Physics, University of Zanjan, PO Box 45195-313, Zanjan, Iran}
\affiliation{Optics Research Center, Institute for Advanced Studies in Basic Sciences, PO Box 45137-66731, Zanjan, Iran.}
 
\author{Er\c{c}a\v{g} Pince}
\affiliation{Soft Matter Lab, Department of Physics, Bilkent University, Cankaya, 06800 Ankara, Turkey}

\author{Gokhan Baris Bagci}
\affiliation{Department of Materials Science and Nanotechnology Engineering, TOBB University of Economics and Technology, 06560 Ankara, Turkey}

\author{Giovanni Volpe}
\affiliation{Soft Matter Lab, Department of Physics, Bilkent University, Cankaya, 06800 Ankara, Turkey}
\affiliation{UNAM -- National Nanotechnology Research Center, Bilkent University, Ankara 06800, Turkey}

\date{\today}

\begin{abstract}
\textbf{
Most natural and engineered processes, such as biomolecular reactions, protein folding, and population dynamics, occur far from equilibrium and, therefore, cannot be treated within the framework of classical equilibrium thermodynamics. The Jarzynski equality holds the promise to calculate the free-energy difference between two states from the Boltzmann-weighted statistics of the irreversible work done along trajectories arbitrarily out of equilibrium. This equality is the subject of intense activity. However, the applicability of the Jarzynski equality to systems far from equilibrium such as living matter has not been investigated yet. We present an experimental test of the Jarzynski equality predictions on a paradigmatic physical model, i.e. a Brownian particle held in an optical potential, coupled either to a thermal bath or to an active bath. While in the thermal bath we find that the Jarzynski equality correctly retrieves the free-energy difference from nonequilibrium measurements, in the active bath the Jarzynski equality fails because of the presence of non-Boltzmann statistics. We corroborate our experimental findings with theoretical arguments and numerical simulations.
}
\end{abstract}

\maketitle

The second law of thermodynamics draws the boundaries between reversible (equilibrium) and irreversible (nonequilibrium) processes. For a macroscopic system driven from state $1$ to state $2$ at absolute temperature $T$, the Clausius version of the second law relates the external work done on the system, $W$, to the free energy difference between the two states, $\Delta F$, as $W \geq \Delta F$ \cite{callen1985thermodynamics} . If the system is driven infinitely slowly so that it always remains at equilibrium with the thermal bath, the process is reversible and the second law holds in its equality form, i.e. $ W  = \Delta F$. Instead, its inequality form, i.e. $W > \Delta F$, holds for irreversible processes \cite{callen1985thermodynamics,evans1993probability,jarzynski1997nonequilibrium,crooks1998nonequilibrium,hummer2001free,reguera2005mesoscopic}. For a microscopic system, the presence of thermal fluctuations allows the violation of the second law for single realizations of the process, but nevertheless the second law still holds on average \cite{wang2002experimental}, i.e. $\left\langle W_i \right\rangle \geq \Delta F$ with $\left\langle W_i \right\rangle = \Delta F$ if the process is infinitely slow, where $W_i$ is the work done during the $i$-th realization of the process and the angular brackets represent the ensemble average. The macroscopic relation is recovered because statistical fluctuations around the average become insignificant and can be dismissed for large systems as a consequence of the law of large numbers \cite{ellis2005entropy}.

A major advance in stochastic thermodynamics has been the realization that, instead of just averaging the fluctuations out, it is possible to make use of the information they store in order to recover equilibrium information even from nonequilibrium measurements \cite{jarzynski2011equalities,seifert2012stochastic}. An important example of such a relation is the Jarzynski equality \cite{jarzynski1997nonequilibrium}, which permits one to obtain the free-energy difference from an exponential average of the work associated with irreversible realizations of a thermodynamic process, i.e.
\begin{equation}\label{eq:je}
\left< e^{-\beta W_i}\right> = e^{-\beta \Delta F},
\end{equation}
\noindent where $\beta = 1/k_{\rm B} T$ is the inverse temperature of the initial equilibrium state of the system and $k_{\rm B}$ is the Boltzmann constant. Until now, only few experimental tests of the Jarzynski equality have been realized: the Jarzynski equality was initially verified by mechanically stretching a single RNA molecule reversibly and irreversibly between two conformations \cite{liphardt2002equilibrium}, while other tests have followed on single molecules \cite{harris2007experimental,berkovich2008analyzing,gupta2011experimental,collin2005verification}, on a mechanical oscillator \cite{douarche2005experimental}, on electronic systems \cite{saira2012test}, on colloidal particles in non-harmonic potentials \cite{blickle2006thermodynamics}, and also on quantum systems \cite{an2015experimental}. However, all these tests considered systems coupled to a thermal bath and, therefore, satisfying Boltzmann statistics. The systems satisfying these conditions are limited and, in particular, do not include many systems that are intrinsically far from equilibrium such as living matter, for which fluctuations are expected to be non-Gaussian \cite{ben2011effective}. For example, biomolecules within the cell are coupled with an active bath due to the presence of molecular motors within the cytoplasm, which leads to striking and largely not yet understood phenomena such as the emergence of anomalous diffusion \cite{barkai2012strange}. Also, protein folding might be facilitated by the presence of active fluctuations \cite{harder2014activity} and active matter dynamics could play a central role in several biological functions \cite{suzuki2015polar,mallory2015anomalous,shin2015facilitation}. It is therefore an open and compelling question to assess whether and to what degree the Jarzynski equality can be applied to systems coupled to active baths.

Here we test the Jarzynski equality using one of the simplest model systems for stochastic thermodynamics \cite{seifert2012stochastic}: a Brownian particle held in a harmonic potential whose position can change over time. Differently from previous works, we consider both a particle coupled to a thermal bath and one coupled to an active bath. As we will see in the following, the Jarzynski equality holds in the thermal bath, but its predictions are incorrect in the case of the active bath. We remark that our results are not in disagreement with previous theoretical and experimental works regarding the Jarzynski equality as these works have not considered the case of active baths.

\begin{figure}[h!]
\includegraphics[width=.5\textwidth]{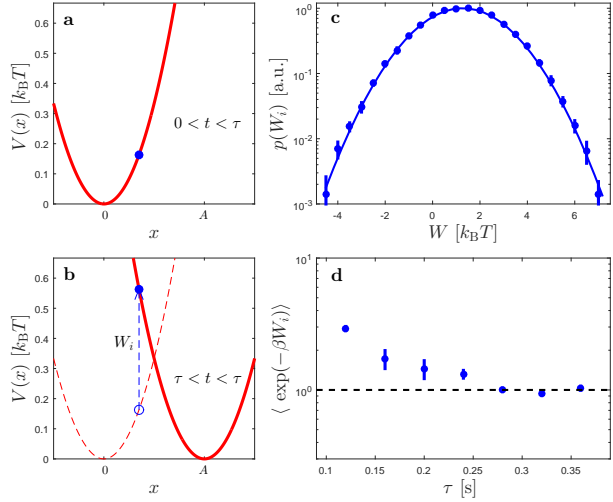}
\caption{{\bf Experimental protocol and verification of the Jarzynski equality in a thermal bath.} 
{\bf a.} A spherical Brownian microparticle (silica, diameter $2R = 4.23\pm0.20\,{\rm \mu m}$ ) is immersed in a thermal bath (watery buffer solution) and held in a harmonic optical potential generated by a focused laser beam. 
{\bf b.} After a delay time $\tau$, the center of the potential is shifted from position $0$ to position $A=90\,{\rm nm}$ and the work done on the particle in the process is evaluated from the measured particle trajectory according to Eq.~(\ref{eq:work}).
{\bf c.} The measured work distribution (symbols) obtained for $\tau = 0.36\,{\rm ms}$ from 5000 experimental realizations of the protocol; it is well-fitted by a Gaussian function (line). 
{\bf d.} As the switching period $\tau$ is increased, the particle has time to thermalize in the potential well before the next switch and the exponential average converges towards 1 (dashed line) as predicted by the Jarzynski equality (Eq.~(\ref{eq:je2})). Each symbol represents the exponential average of 5000 experimental realizations of the protocol. The error bars represent standard errors.}
\label{figure1}
\end{figure}

We start by considering the case of a spherical Brownian microparticle (silica, diameter $2R = 4.23\pm 0.20\,{\rm \mu m}$) in a thermal bath at absolute temperature $T$ trapped in a harmonic optical trap generated by focusing a laser beam (wavelength $\lambda = 532\,{\rm nm}$, power $2.1\,{\rm mW}$) with a high-numerical-aperture objective (oil-immersion, $\times 100$, ${\rm NA} = 1.30$) \cite{jones2015optical}. The particle is held by the optical trap restoring force, while also thermal forces and viscous drag forces act. The particle position is monitored by digital video microscopy with a spatial resolution of $5\,{\rm nm}$ and a sampling frequency of $50$ frames per second. A more detailed description of the setup is provided in the Methods and in Supplementary Fig.~\ref{figure-S1}. The resulting trapping potential is 
\begin{equation}\label{pot}
V\left(x, t\right) = {1 \over 2} k \left( x-\lambda(t)\right)^{2}  \; ,
\end{equation}
\noindent where $x$ is the position of the particle, $t$ is time, $k = 1.33 \pm 0.05 \,{\rm pN/\mu m}$ is the trap stiffness, and $\lambda (t)$ denotes the center of the harmonic potential, which acts as our external control parameter. In our protocol, $\lambda(t)$ is a square-wave function with amplitude $A$ and period $2\tau$: first, the potential is centered at position $0$ (state 1, Fig.~\ref{figure1}a) for a time interval $\tau$; then, it is shifted to position $A=90\,{\rm nm}$ (state 2, Fig.~\ref{figure1}b), where it remains for an additional time interval $\tau$; finally, the potential is brought back to position $0$ and the protocol is iterated. The free energy of the optically trapped particle is 
\begin{equation}\label{free}
F = - {1\over\beta} \ln \left[ \int dx \; e^{-\beta V(x)} \right]  = -\frac{1}{2\beta} \ln \left( \frac{2 \pi}{k} \right) \; ,
\end{equation}
\noindent which depends only on $k$ and, importantly, not on the trap center position \cite{abreu2011extracting}. Thus, since in our protocol the stiffness is kept constant, the free energy difference between state 1 and state 2 is zero, i.e. $\Delta F = 0$, and the Jarzynski equality (Eq.~(\ref{eq:je})) simplifies to
\begin{equation}\label{eq:je2}
\left< e^{-\beta W_i}\right> = 1 \; ,
\end{equation}
\noindent where the work-weighted propagator in the path integral approach allows an exact analytical result for this particular model \cite{adib2009path}. In close scrutiny, it can be seen that the validity of the Jarzynski equality for this particular model stems from the fact that the initial distribution is Boltzmann \cite{adib2009path}, as detailed in the Methods. 

We have experimentally verified the validity of Eq.~(\ref{eq:je2}), and therefore of the Jarzynski equality, for our experimental system. We repeated the experimental protocol described above 5000 times; for each repetition, we recorded the particle trajectory $x_i(t)$ and evaluated the work done on the particle according to the path integral
\begin{equation}\label{eq:work}
W_i \left[ x_i(t)\right] = \int_{0}^{\tau} dt \; \frac{\partial V(x_i(t),t)}{\partial t} \; . 
\end{equation}
The resulting experimental work distribution is shown by the symbols in Fig.~\ref{figure1}c. We then evaluated the exponential average of the work, i.e. $\left< e^{-\beta W_i}\right>$: as the switching time $\tau$ is increased, the particle is given time to thermalize within the potential well before the potential is switched, and $\left< e^{-\beta W_i}\right>$ converges towards 1, as shown by the symbols in Fig.~\ref{figure1}d, in agreement with the prediction of the Jarzynski equality (dashed line in Fig.~\ref{figure1}d, Eq.~(\ref{eq:je2})).

\begin{figure}[h!]
\includegraphics[width=.5\textwidth]{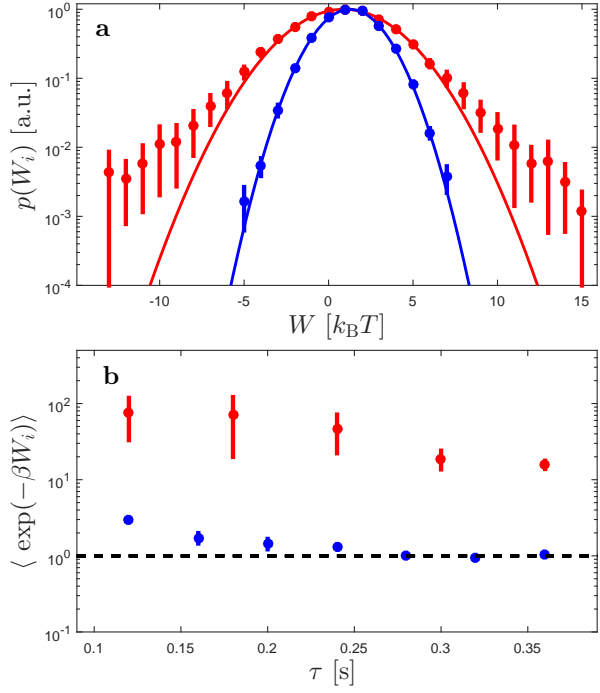}
\caption{{\bf Failure of the Jarzynski equality in an active bath.}
{\bf a.} In an active bath, the measured work distribution (red symbols) is not Gaussian, as can be seen from the deviations from the best Gaussian fit (red line); in particular, the work distribution is heavy-tailed. The blue symbols and line reproduce the (Gaussian) work distribution for the thermal  bath case reported in Fig.~\ref{figure1}c.
{\bf b.} The red symbols show the exponential average of the work done on the particle in the active bath; they feature a clear deviation from the value expected according to the Jarzynski equality (dashed line, Eq.~(\ref{eq:je2})); differently form the thermal bath case (see Fig.~\ref{figure1}d, the corresponding data are reproduced here by the blue symbols for comparison); this deviation does not decrease as the switching period $\tau$ is increased. The error bars represent standard errors.}
\label{figure2}
\end{figure}

In the second part of our experiment, we repeat the same procedure and analysis, but considering a particle in an active bath, instead of a passive bath. We realize the active bath by adding motile bacteria to the watery solution where the particle is immersed \cite{wu2000particle}: the bacteria behave as active particles and exert non-thermal forces on our probe particle so that it experiences non-thermal fluctuations and features a qualitatively different behavior from that of a Brownian particle in a thermal bath. We proceed to test the Jarzynski equality with the same protocol employed above for the thermal bath case. First, we repeat the experimental protocol 5000 times and measure the distribution of the work done on the particle when switching back and forth between state 1 and state 2; the experimental results are shown by the red symbols in Fig.~\ref{figure2}a. Then, we use these data to evaluate the exponential averages of the work done on the particle as a function of $\tau$, which are shown by the red symbols in Fig.~\ref{figure2}b: differently from the thermal bath case (blue symbols in Fig.~\ref{figure2}b), they do not converge towards the prediction of the Jarzynski equality. Therefore, we conclude that in the presence of an active bath the Jarzynski equality fails and cannot be used to obtain free energy differences from out-of-equilibrium measurements. We remark that, differently from the case in the passive bath, in an active bath the work distribution is not Gaussian but heavy-tailed, as shown in Fig.~\ref{figure2}a, where the experimental data for the work distribution (red symbols) are compared to the best Gaussian fit (red line), which shows clear deviations in the tails (heavy tails).

\begin{figure}
\includegraphics[width=.5\textwidth]{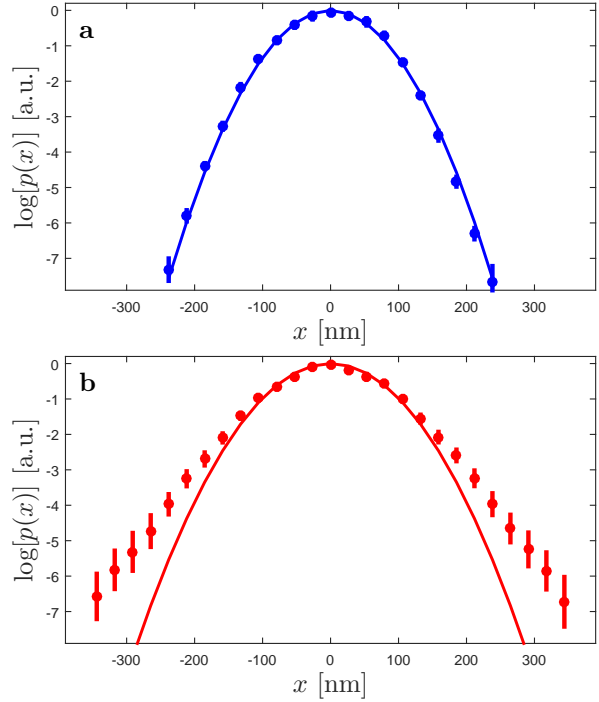}
\caption{{\bf Non-Boltzmann statistics in an active bath.} {\bf a.} The experimental probability distribution of the position of a particle (symbols) held in a harmonic potential in a thermal bath is Gaussian (line, corresponding to the theoretical prediction for a particle held in a harmonic potential with $k = 1.11 \pm 0.07 \,{\rm pN/\mu m}$). This corresponds to the Boltzmann statistics associated with the harmonic potential. {\bf b.} However, the experimental position distribution (symbols) in an active bath deviates from a Gaussian (best fit represented by the line) and features heavy tails. This is a signature that the statistics in the active bath are non-Boltzmann and that the behavior of a particle in an active bath is qualitatively different from that in a thermal bath, i.e. it cannot correspond to a different (higher) effective temperature. The data are obtained from $200\,000$ position measurements.}
\label{figure3}
\end{figure}

In order to understand more deeply the difference between the thermal bath and the active bath cases and, thus, the underlying reason why the Jarzynski equality fails in active baths, we also compare the probability distribution of the position of a particle held in a harmonic optical potential in the two cases. To this end, we acquire the particle position in both cases while keeping the optical trap fixed. The results are shown in Fig.~\ref{figure3}. In the thermal bath (Fig.~\ref{figure3}a), the experimental distribution of the particle position (symbols) is Gaussian (line), as expected for the Boltzmann distribution of a Brownian particle in a harmonic trap \cite{jones2015optical}. In the active bath instead (Fig.~\ref{figure3}b), the experimental position probability distribution (symbols) is non-Gaussian, as can be seen by comparing these data with the best Gaussian fit (line). This becomes particularly evident when considering the tails of the distribution, which are clearly heavier than expected from the Gaussian distribution. Importantly, the distribution is non-Boltzmann, which implies that the case of a particle in an active bath is qualitatively different from the case of a particle in a thermal bath. In fact, it is not possible to recover this probability distribution by altering the effective temperature of the system, as the position probability distribution associated to a Brownian particle in a harmonic potential is Gaussian at all temperatures. The alteration of the distribution is in fact due to the correlations introduced in the motion of the particle by the active bath, which lead to a non-Gaussian noise. 

We have further supported our conclusions with numerical simulations, which are described in the Methods and whose results are shown in Supplementary Fig.~\ref{figure-S2}. The numerical results are in good agreement with the experimental results we have presented. On the one hand, also the simulations show that, in a thermal bath, the work distribution is Gaussian (Supplementary Fig.~\ref{figure-S2}a), the Jarzynski equality is verified as $\tau$ grows (Supplementary Fig.~\ref{figure-S2}b), and the particle position distribution within a static harmonic potential is given by the Boltzmann distribution (Supplementary Fig.~\ref{figure-S2}c). On the other hand, in an active bath, the work distribution becomes non-Gaussian (Supplementary Fig.~\ref{figure-S2}d), the Jarzynski equality fails even as $\tau$ grows (Supplementary Fig.~\ref{figure-S2}e), and the particle position distribution within a static harmonic potential is non-Boltzmann and it is not possible to interpret it as a different (higher) effective temperature (Supplementary Fig.~\ref{figure-S2}f).

In conclusion, we have shown that in an active bath the statistical properties of a particle held in a potential do not follow the Boltzmann statistics; this is a consequence of the noise with long-time correlation introduced by the active bath and cannot be modelled by introducing an effective temperature. More importantly, we have also shown that a major consequence of this fact is that Jarzynski equality, as stated in its classical formulation, fails in active baths. Since active matter plays a central role in many systems, including very importantly living systems, our findings pose some significant limitations to the possibility of applying nonequilibrium fluctuation-dissipation relations, such as the Jarzynski equality, to study this broad and valuable class of systems, pointing to the need for alternative approaches that explicitly model the presence of non-thermal fluctuations.

\section*{Methods}

\subsection{Experimental setup}

\renewcommand{\figurename}{{\bf Supplementary Figure}}
\setcounter{figure}{0}

\begin{figure}[h!]
\begin{sideways}
\includegraphics[width=.5\textwidth]{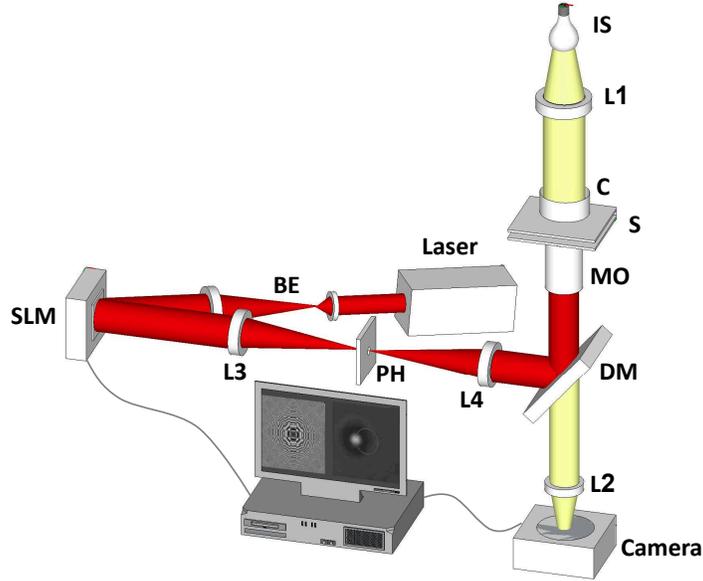}
\end{sideways}
\caption{{\bf Schematic of the experimental setup.}  The optical setup is based on a holographic optical tweezers and a digital video microscope. IS: Illumination source. L1: Collecting lens. C: Condenser. S: Sample chamber. MO: Microscope objective. DM: Dichroic mirror. L2: Tube lens. BE: Beam expander. SLM: Spatial light modulator. L3 and L4: lenses for 4f-configuration. PH: Pinhole. }
\label{figure-S1}
\end{figure}

The experimental setup consists of a holographic optical tweezers system built on a homemade optical microscope and integrated with a digital video microscopy system \cite{jones2015optical}. A schematic of the setup is shown in Supplementary Fig.~\ref{figure-S1}. A laser beam (Coherent Verdi-V6, wavelength $\lambda = 532\,{\rm nm}$) is expanded by the beam expander BE and projected onto a reflective phase-only spatial light modulator (SLM, Holoeye LC-R 720). The SLM is placed in a plane conjugated to the back aperture of the microscope objective (MO) using a 4f-configuration (lenses L3 and L4). The first-order diffracted beam is spatially selected by a pinhole (PH, at the focal point between lenses L3 and L4) and directed onto the high-numerical-aperture microscope objective (MO, Nikon $100\times$, oil immersion, numerical aperture ${\rm NA}=1.30$, working distance ${\rm WD}=0.17$) by a dichroic mirror (DM). The SLM permits us to shape the phase of the incoming beam and, therefore, the profile of the optical potential generated by the focused beam. The optical power of the first-order diffracted beam at the back-focal plane of the trapping objective is kept constant at a value of $2.1\,{\rm mW}$ for the experiments reported in Figs.~\ref{figure1} and \ref{figure2}, and $1.8\,{\rm mW}$ for those reported in Fig.~\ref{figure3} in order to maintain the stiffness of the optical trap constant within each experimental set. The imaging system (IS) consists of a quasi-monochromatic light emission diode, a collecting lens (L1), a $20\times$ microscope objective working as condenser (C), the microscope objective (MO), a tube lens (L2), and the camera. We record videos of the optically trapped particles using a CMOS camera (Thorlabs DCC1645C-HQ) adjusting the illumination and the tube lens so that the particle looks as a bright spot with a maximum of intensity at the center. We track the particle position using the center of mass algorithm \cite{jones2015optical} obtaining a spatial resolution of $5\,{\rm nm}$. We have calibrated the optical potential using standard calibration algorithms \cite{jones2015optical}, which permit us to verify that the potential is harmonic and to measure the trap stiffness $k=1.33 \pm 0.05 \,{\rm pN/\mu m}$ for the experiments reported in Figs.~\ref{figure1} and \ref{figure2}, and $k=1.11 \pm 0.07 \,{\rm pN/\mu m}$ for those reported in Fig.~\ref{figure3} \cite{jones2015optical}.

\subsection{Bacterial bath preparation}

\emph{Escherichia coli} bacteria were cultured from the wild-type strain RP437 provided by the \emph{E. coli} Stock Center at Yale University. The liquid culture of RP437 was taken from a $-80^{\circ}{\rm C}$ archive stock and streaked onto a sterile hard agar medium. The inoculated agar plate was incubated and grown overnight at $32.5^{\circ}{\rm C}$. Single colonies grown on agar were isolated by sterile toothpicking and inoculated into a fresh liquid growth medium containing tryptone broth ($\%$1 tryptone). After reaching the saturation phase, the culture was diluted 1:100 into a fresh tryptone broth. The final dilution was incubated at $32.5^{\circ}{\rm C}$ and mildly shaken at $180\,{\rm rpm}$ until the culture reached its middle growth phase (OD $600\sim0.40$). Then, $7\,{\rm ml}$ of the final dilution were transferred into a falcon tube and centrifuged at $2000\,{\rm rpm}$ at room temperature for 10 minutes. Precipitated RP437 pellets were then gently collected and immersed in $4\,{\rm mL}$ of motility buffer containing $10\,{\rm mM}$ monobasic potassium phosphate (${\rm KH_2PO_4}$), $0.1\,{\rm mM}$ EDTA, $10\,{\rm mM}$ dextrose, and $0.002\%$ of Tween 20. The cell collection and resuspension procedure was repeated three times in order to undermine the growth medium and terminate further cell duplication inside the motility buffer.

\subsection{Jarzynski equality for a Brownian particle in a moving harmonic potential}

\begin{figure}[b!]
\includegraphics[width=.5\textwidth]{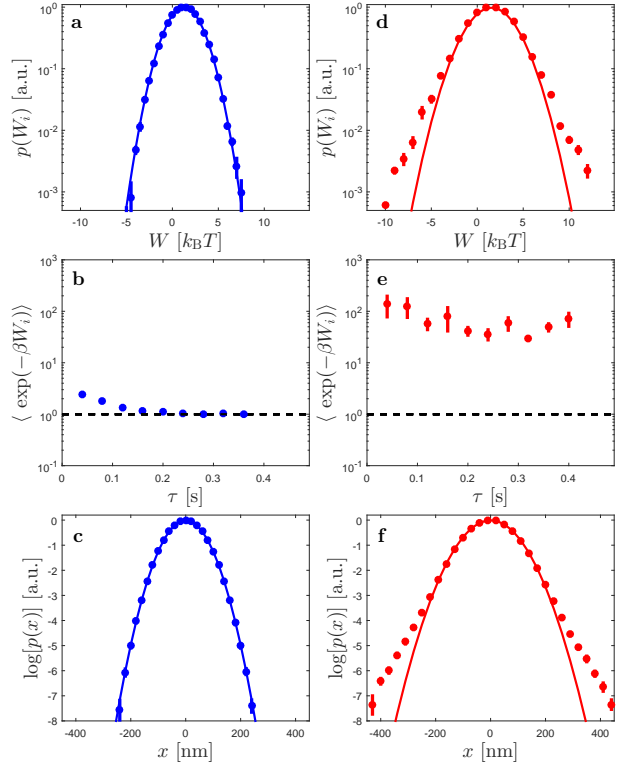}
\caption{{\bf Numerical results.} {\bf a}, {\bf b} and {\bf c} represent the numerical results obtained for the work distribution, the verification of the Jarzynski equality (Eq.~(\ref{eq:je2})), and the probability distributions inside a fixed optical trap, respectively, in a thermal bath. {\bf d}, {\bf e} and {\bf f} represent the corresponding numerical results in an active bath, where Jarzynski equality (Eq.~(\ref{eq:je2})) fails. There is good agreement between these numerical results and the experimental data presented in Figs.~\ref{figure1}, \ref{figure2}, and \ref{figure3}.}
\label{figure-S2}
\end{figure}

The Jarzynski equality for a Brownian particle coupled to a thermal bath and held in a moving harmonic potential with constant stiffness (Eq.~(\ref{eq:je2})) can be analytically verified by the path integral method \cite{adib2009path}. The integration is performed by considering a  work-weighted function  over all possible trajectories between the initial  and final positions of the particle. The work-weighted propagator for a time-dependent  potential such as $V(x,t)=k(x-vt)^2/2$ is found to be
\begin{equation}
p_w=\frac{\exp \left( -\frac{k}{2} \frac{(x(t)-vt-\frac{v}{Dk}-(x_0-\frac{v}{Dk})e^{-Dk \tau_{\rm s}})^2}{1-e^{-2Dk \tau_{\rm s}}}+ \frac{v^2t}{D} \right) }{\sqrt{ \frac{2 \pi}{k} (1-e^{-2Dk \tau_{\rm s}})}} \; ,
\end{equation}
where $x(t)$ denotes the position of the particle, $x_0$ is its initial position, D is its diffusion coefficient, $v$ is velocity at which the center of the harmonic potential is shifted, and $ \tau_{\rm s}$ is time interval during which this shift occurs. The canonical integral over all possible initial and final positions of the particle leads to the Jarzynski average as
\begin{equation}
\langle e^{-w} \rangle = \int dx_0 \int dx \frac{e^{-V(x_0,0)}}{\int dy e^{-V(y,0)}} p_w(x,\tau_{\rm s} | x_0,0) = 1 \; ,
\end{equation}
which is Eq.~(\ref{eq:je2}).
The validity of this equality hinges on the fact that the initial particle position is distributed according to the Boltzmann distribution, which in the case of a harmonic potential is a Gaussian distribution.

\subsection{Numerical simulations}

The data presented in Supplementary Fig.~\ref{figure-S2} show the analysis of trajectories obtained from Brownian dynamics simulations of an optically trapped particle in a thermal bath and in an active bath. In a thermal bath, the forces acting on the particle are the optical trap restoring force $-kx$, a viscous drag force $-\gamma \dot{x}$, and a thermal noise $\sqrt{2\gamma k_{\rm B} T}\xi(t)$, where $ \gamma$ is the friction coefficient of the particle and $\xi(t)$ is a Gaussian white process with zero mean and unitary variance. We simulate the overdamped Langevin equation corresponding to the optically trapped particle in a thermal bath following the procedures explained in Ref.~\cite{volpe2013simulation} and using the experimental parameters. In order to simulate the particle's motion in an active bath, we added an extra term to the equation to account for the bacteria pushing the trapped particle, which make the particle behave as an active particle \cite{wu2000particle}. 

\bibliography{biblio}

\textbf{Funding}
ARM has been partially supported by  Scientific and Technological Research Council of Turkey (TUBITAK) under Grant No. 114F207.
GV has been partially supported by Marie Curie Career Integration Grant (MC-CIG) PCIG11 GA-2012-321726 and a Distinguished Young Scientist award of the Turkish Academy of Sciences (T\"UBA).

\textbf{Contributions} 
GV conceived and supervised the project. AA and ARM designed and carried out experiments. AA, ARM, and GBB  analyzed and interpreted the data. GBB provided theoretical support. AA implemented the simulations.  EP prepared the bacterial samples. All authors contributed to the manuscript preparation. 

\textbf{Competing Interests}
The authors declare that they have no competing financial interests.

\textbf{Correspondence} 
Correspondence and requests for materials should be addressed to Giovanni Volpe~(email: giovanni.volpe@fen.bilkent.edu.tr).

\end{document}